\documentclass[dvips]{article}
\usepackage{icrctc07}
\title{Blazar Observations with VERITAS}
\shorttitle{VERITAS Blazar Observations}
\authors{H. Krawczynski$^{1}$ for the VERITAS Collaboration$^{2}$.}
\shortauthors{H. Krawczynski and et al}
\afiliations{$^1$ Washington University in St. Louis, Physics Department, 1 Brookings Dr., CB 1105, St. Louis, MO 63130}
\afiliations{$^2$ For full author list see G. Maier, ''Status and Performance of VERITAS'',
these proceedings.}
\email{krawcz@wuphys.wustl.edu}

\abstract{The Very Energetic Radiation Imaging Telescope Array System (VERITAS) is an array of four 12m diameter Imaging Atmospheric Cherenkov Technique (IACT) telescopes operated at the base of Mt. Hopkins in southern Arizona. The four-telescope experiment started operation in April, 2007. GeV and TeV gamma-ray observations of blazars can be used to probe the structure and composition of their jets, and to contribute to our understanding of how supermassive black holes accrete matter. In this contribution, we present first VERITAS blazar results obtained with three and four telescopes.}
\begin{document}
\maketitle
\section{Introduction}
The EGRET (Energetic Gamma Ray Experiment Telescope) detector on board 
the Compton Gamma-Ray Observatory discovered strong MeV $\gamma$-ray
emission from 66 Active Galactic Nuclei (AGNs), mainly from 
Flat Spectrum Radio Quasars and Flat Spectrum Radio Sources \cite{1999yCat..21230079H}. 
As of writing these proceedings in May 2007, ground-based Cherenkov telescopes have
discovered TeV $\gamma$-ray emission from 17 AGB \cite{2004NewAR..48..367K}. 
Sixteen of the 17 sources are blazars and one is the radio galaxy M 87 \cite{2006Sci...314.1424A,2003A&A...403L...1A}.
The blazars are mainly high energy peaked BL Lac objects, with BL Lac itself (an
intermediate peaked BL Lac) being the only exception \cite{Albe:07}.
The redshifts of the TeV $\gamma$-ray sources range from $z\,=$ 0.031 for Mrk 421 
\cite{1992Natur.358..477P} to $z\,=$ 0.188 for 1ES 0347-121 \cite{2006Natur.440.1018A}.

In this contribution, we will give an overview of the blazar observations
performed with the VERITAS experiment. VERITAS is an array of four
12~m diameter Cherenkov telescopes located at an altitude of 1268 m above sea level
on Mt. Hopkins, Az (31$^{\circ}$ 40' 30.21" N, 110$^{\circ}$ 57' 07.77" W) 
\cite{2002APh....17..221W}. The experiment started 
operation with two telescopes in spring 2006, and with four telescopes 
in winter 2006. The telescope system achieves an angular resolution of 
0.16$^{\circ}$ and a 250 GeV-1 TeV $\nu F_{\nu}$ sensitivity of 
10$^{-12}$ ergs cm$^{-2}$ s$^{-1}$ for 10 hours of integration.
For a detailed description of the status and performance of the telescope system the 
reader is referred to the contributions of Meier et al. \cite{Meie:07} and Celik et al. \cite{Celi:07} in this volume.
The most important blazar detections are described in dedicated contributions,
please see Fortin et al. \cite{Fort:07} for the 1ES 1218+304 results,
Cogan et al. \cite{Coga:07} for the 1ES 0806+524 and 1ES 647+250 result, 
Fegan et al. for Mrk 421 and Mrk 501 results \cite{Fega:07}, and 
Colin et al. \cite{Coli:07} for the M 87 results.
The Whipple 10 m Cherenkov telescope is used to monitor blazars on a regular basis.
The results of the observations taken in 2006 are described by Steele et al. \cite{Stee:07}.
\begin{table}[t]
\begin{center}
\begin{tabular}{p{1.4cm}p{1.4cm}p{3.1cm}}
\hline
Observatory 	& Wavelength 	& Contact\\ \hline
{ Owen V.}	& Radio 	& A. Readhead\\
Metsahovi 	& Radio 	& A. Lahteenmaki\\
WEBT		& Radio/IR/O 	& M. Villata\\
Abastumani	& Opt. 		& O. Kurtanidze\\
Antipodal	& Opt.  	& J. Buckley\\
Bell 		& Opt.  	& M. Carini\\
Boltwood  	& Opt.  	& P. Boltwood\\
Bordeaux	& Opt.	     	& P. Charlot\\
Tuorla   	& Opt. 		& A. Sillanpaa\\
WIYN 0.9m  & Opt.          & T. Montaruli\\
Swift		& X-ray 	& H. Krimm\\
AGILE		& $\gamma$-ray 	& M. Tavani\\
GLAST		& $\gamma$-ray 	& J. McEnery\\
MAGIC		& $\gamma$-ray  & D. Mazin\\
H.E.S.S.	& $\gamma$-ray	& S. Wagner\\
IceCUBE		& Neutrino 	& T. Montaruli\\
\hline
\end{tabular}
 \caption{List of VERITAS multiwavelength collaborators. 
Only one contact person is given for each observatory.}
\end{center}
\end{table}
\section{The VERITAS Blazar Observation Program and Multiwavelength Coverage}
Blazar observations are one of four VERITAS key science projects that will
be performed during the first two years of the operation of the four telescope system.
The other key science projects concern a scan of the galactic plane, supernova 
remnant observations, and the search for $\gamma$-rays from dark matter annihilation.
The blazar key science project receives 115 hrs of observation time per year;
all four key science projects receive 400 hrs, or 50\% of the observation time.
The key science project was developed by the blazar science working group and 
includes three equally observation-intensive components: (i) multiwavelength 
observations of bright blazars in flaring state with dense coverage, (ii) deep 
observations of a small number of six blazars at a 
range of redshifts to obtain high-quality energy spectra, and
(iii) the search for emission from different types of blazars including
high energy peaked BL Lac objects (HBLs), intermediate energy peaked BL Lac 
objects (IBLs), and flat spectrum radio quasars (FSRQs).
The first two components aim at identifying the emission process, constraining the
jet composition and structure and thus the process of jet formation, and detecting
the imprint of absorption owing to pair creation processes of 
TeV $\gamma$-rays interacting with infrared photons of the extragalactic 
background light. The aim of the third component is to explore the 
$\gamma$-ray emission characteristics in different types of blazars.

The first and second component depend on the detection of a blazar flaring state.
We use the Whipple 10m Cherenkov telescope, the third most sensitive ground-based
$\gamma$-ray telescope in the northern hemisphere, to monitor the sources of interest.
The Whipple 10~m telescope can detect flares on the level of 70\% of the flux from 
the Crab Nebula within one hour. In addition, as described below, VERITAS respond 
to alerts from other Cherenkov telescope experiments.

Excellent multiwavelength coverage is key to achieving the science objectives.
The VERITAS collaboration has established collaborations with the
observatories listed in Table 1. The collaborative activities include the 
planning of the VERITAS blazar observation program and the joint publication of 
observational results. In case of the two $\gamma$-ray observatories MAGIC and
H.E.S.S., an agreement was reached to alert each other about noteworthy flares
of all well established sources of $\gamma$-rays as soon as they have been 
detected. Furthermore, close collaboration in multiwavelength campaigns is 
envisioned.
\section{Results}
VERITAS has detected the sources Mrk 421, Mrk 501, and 1ES 1218+304.
All these sources plus 1ES 0806+524 and 1ES 0647+250 are described in
more detail in dedicated ICRC contributions (see Table 2).

In Table 2, we show other blazars that VERITAS has observed so far. 
The data have been analyzed using independent analysis packages \cite{Dani:07}. 
All of these analyses yield consistent results.
Only runs with a cosmic ray rate (corrected for the zenith angle dependence)
deviating by less than 20\% from the average rate have been used for the analysis.
The most important event selection cuts are mean scaled width and mean scaled 
length parameters smaller than 0.5, and an angular deviation of smaller than 0.158$^{\circ}$
from the nominal source position. After analysis cuts, the peak energy is about 250 GeV.
The peak energy is the energy at which the differential detection rate peaks 
for a Crab like energy spectrum. The reflected region background model is used for
background estimation with an on-off solid angle ratio of 1:4 \cite{2007A&A...466.1219B}. 
We calculate significances with the equation 17 of \cite{1983ApJ...272..317L}. 
\begin{table}[t]
\begin{center}
\begin{tabular}{p{1.7cm}p{0.6cm}p{0.8cm}p{0.8cm}p{1cm}}
\hline
Source 		 	& $z$	& Time 		& Sign. 	& Ref. \\
 			&	& [hours] 	& [$\sigma$]	&      \\ \hline
{\footnotesize \hspace*{-.4cm} Mrk 421}			& 0.031	& 4.5 & 35 &\cite{Fega:07} \\
{\footnotesize \hspace*{-.4cm} Mrk 501} 		      & 0.034	& 12.5 & 16 &\cite{Fega:07} \\
{\footnotesize \hspace*{-.4cm} 1ES 1218+304} 		& 0.138	& 17.4 & 10.2 &\cite{Fort:07} \\
{\footnotesize \hspace*{-.4cm} M87 }		 	& 0.004	& 44.2 & 5.1  &\cite{Coli:07} \\
\hline
\end{tabular}
 \caption{Blazars observations performed with VERITAS that are described in detail in other ICRC contributions.
All sources are High-Energy Peaked BL Lac Objects, except M87 which is a radio galaxy.}
\end{center}
\end{table}

\begin{table}[t]
\begin{center}
\begin{tabular}{p{1.8cm}p{0.6cm}p{0.8cm}p{0.8cm}p{1cm}}
\hline
Source 		& $z$	& Time 		& Sign. 	& UL \\
 		&	& [hours] 	& [$\sigma$]	&{\footnotesize [\%Crab]}\\ \hline
\multicolumn{5}{c}{HBL}\\ \hline
{\footnotesize \hspace*{-.4cm} 1ES 1011+496} 	& 0.200 &0.67 & 2.1 & 8.6\\
%
%
{\footnotesize \hspace*{-.4cm} RXJ1211+2242}	& 0.455 &1&-1.5&2.0\\
%
{\footnotesize \hspace*{-.4cm} H1426+428   }	& 0.129 &12.5&3.2&2.9\\
%
\hline \multicolumn{5}{c}{FSRQ}\\ \hline
{\footnotesize \hspace*{-.4cm} 3C279       }	& 0.536 &2&0.7&4.7\\ 
{\footnotesize \hspace*{-.4cm} RGBJ1413+436}	& 0.090 &2.7&0.25&2.8\\ 
{\footnotesize \hspace*{-.4cm} 1ES 1627+402 }	& 0.271 &10.1&1.5&2.2 \\[-0.25ex] 
\hline
\end{tabular}
 \caption{List of some of the blazars observed by VERITAS not listed in Table 2. Flux upper limits are 
given on 99\% confidence level in units of the flux from the Crab Nebula.
}
\end{center}
\end{table}

This BL Lac object H1426+428 is a well established and well studied source of GeV/TeV $\gamma$-rays
\cite{2002ApJ...571..753H,2002ApJ...580..104P,2002A&A...391L..25D,2003A&A...403..523A}.
When the source was first detected, a GeV/TeV-flux of $\sim$20\% of the flux from the Crab Nebula
was reported \cite{2002ApJ...571..753H}. The analysis reveals a marginal excess (see Fig. 1). 
For this source and all other sources, Bayesian upper limits on the number 
of source counts are derived for a confidence 
interval of 99\% following the recipe described in \cite{Helene:1982pb}. The upper limits 
on the number of source counts are converted to upper limits in ''Crab units'' making 
use of January and February observations of the Crab Nebula.
The upper limits lie in the range from 2.2\% to 8.6\% of the Crab flux.
\begin{figure}
\begin{center}
\vspace*{2ex}
\includegraphics [width=0.48\textwidth]{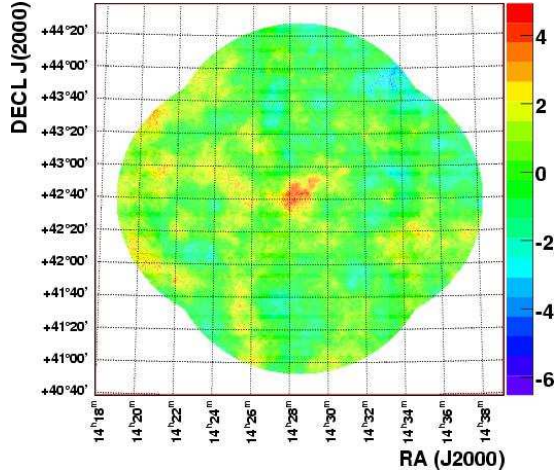}
\end{center}
\caption{Point source significance map from a sky region around the direction of the BL Lac object H1426+428. 
A marginally significant excess can be recognized at the nominal position of the source. The 
statistical significance of the excess is 3.2~$\sigma$.
}\label{fig1}
\end{figure}
\section{Summary}
The VERITAS AGN program is fully underway. The program includes intensive
multiwavelength observations of blazars in a flaring state, deep observations
of blazars to determine their energy spectra with high accuracy, and
the search for TeV $\gamma$-ray emission from a wide range of different types of blazars.
The VERITAS collaboration is working together with a large number of observers
to sample the spectral energy distribution of blazars along the entire electromagnetic 
energy spectrum, and to obtain complementary information through the detection 
of high-energy neutrinos. Agreements have been reached to assure a fruitful collaboration 
between the three Cherenkov telescope experiments VERITAS, MAGIC, and H.E.S.S..

The first observations have resulted in the highly significant detection of the 
blazars Mrk 421, Mrk 501, 1ES 1218+304, and M 87. In this contribution, we have described
3-telescope observations of a number of HBL, IBL, and FSRQs. The flux upper limits are
between 2.2\% and 8.6\% of the flux from the Crab Nebula.
We anticipate exciting results with the full VERITAS system of 4 telescopes.
\section{Acknowledgments}
VERITAS is supported by grants from the U.S. Department of Energy, the
U.S. National Science Foundation and the Smithsonian Institution, by NSERC in
Canada, by PPARC in the U.K. and by Science Foundation Ireland.

\bibliography{blazar01}
\bibliographystyle{prsty}
\end{document}